# Conductance Suppression due to Correlated Electron Transport in Coupled Double-dots

Géza Tóth[*], Alexei O. Orlov, Islamshah Amlani, Craig S. Lent, Gary H. Bernstein, and Gregory L. Snider

*Department of Electrical Engineering*
*University of Notre Dame, Notre Dame, IN 46556*
*e-mail: Geza.Toth.17@nd.edu, Craig.S.Lent.1@nd.edu*

[*]*Also Neuromorphic Information Technology Graduate Center*
*Kende-u. 13, Budapest H-1111, Hungary*

## ABSTRACT

The electrostatic interaction between two capacitively-coupled metal double-dots is studied at low temperatures. Experiments show that when the Coulomb blockade is lifted by applying appropriate gate biases to both double-dots, the conductance through each double-dot becomes significantly lower than when only one double-dot is conducting. A master equation is derived for the system and the results obtained agree well with the experimental data. The model suggests that the conductance lowering in each double-dot is caused by a single-electron tunneling in the other double-dot. Here, each double-dot responds to the instantaneous, rather than average, potentials on the other double-dot. This leads to correlated electron motion within the system, where the position of a single electron in one double-dot controls the tunneling rate through the other double-dot.

# I. Introduction

In the last decade much attention has been given to single charge tunneling phenomena.[1-25] Various aspects have been studied: single dot[13] and double-dot experiments,[18,20,38] single electron transistors,[1-4] single electron turnstile[14] and pump.[19] Both first[5] and second order[6-7,15-16] tunneling phenomena have been analyzed. Correlated transport has also been discussed in the literature. Refs. 9, 12 and 21 analyze the transport of electron-hole pairs (excitons) through arrays of capacitively-coupled double-dots.[26]

This paper is based on a recent experiment realizing a single Quantum-dot Cellular Automata (QCA) cell.[36-37] Although the physical phenomenon to be described is a general feature of coupled double-dots, we review this topic briefly. A QCA cell consists of four metal islands (dots) as shown in Fig.1(a). (In addition to the metal-island cell, the semiconductor quantum-dot and molecular realizations were also studied.[27-35]) The lines in the diagram indicate the possibility of interdot tunneling. The cell has two allowed charge polarizations, P=+1 and -1, as the two extra electrons occupy antipodal sites (Fig. 1(b)). When placed in close proximity along a line, QCA cells align with the same polarization.

The four metal (aluminum) dot system used in this experiment can be seen in Fig. 2(a). The voltage sources, $V_{Dleft}$ and $V_{Dright}$, apply small biases, and currents $I_{left}$ and $I_{right}$ are measured. A symbolic representation of the four dots is shown in Fig. 2(b). The circles denote the dots, and the lines indicate the possibility of interdot tunneling. $D_1$ and $D_2$ are the left double-dot(DD); $D_3$ and $D_4$ are the right DD.

In measuring the conductance through one double-dot (DD) a significant (35-40%) conductance lowering was observed if the other DD was also conducting. This will be referred to as *conductance suppression* in this paper. Our analysis reveals that the cause of the conductance suppression is correlated electron transport in the whole two-DD system; that is, one DD responds to the instantaneous position of the electron in the other DD, and not to the average potential caused by the alternation of the charge configurations in the other DD. In the latter case, the conductance lowering would not happen.

In Sec. II the experiments are explained in detail. In Sec. III the theoretical model is described. In Sec. IV the experimental results and those obtained from the model are compared. The Appendix gives some details about the computation of the average P=+1/P=-1 transition time.

## II. Experiment

In the experiment (for details see Ref. 37) we considered the behavior of a QCA cell, consisting of the two double-dots, to determine the best conditions for QCA operation. The signs of the gate biases were chosen to allow movement of an electron within a double-dot while keeping the total number of electrons constant. We noticed that conductance decreased in both DDs whenever both were conducting.

To understand the experiment we need to examine the charging processes of a two-DD system. The behavior of one DD can be described by the so-called *honeycomb*[1-3,18] graph. This is a phase diagram giving the minimum energy charge configurations as the function of the two electrode voltages. For the whole two-DD system, the electrode voltages of both DDs must also be included in the full description; however, this would mean that the ground state charge configuration must be given as a function of four parameters. In our experiment symmetric input voltages were applied for the DDs. This reduces the number of parameters to two and the occupancy can now be given as a function of $V_{left}=V_{left1}=-V_{left2}$ and $V_{right}=V_{right1}=-V_{right2}$.

Fig. 3(a) shows the phase diagram of the two-DD system if there is no capacitive coupling between the left and right DD. The phases corresponding to different minimum energy charge configurations are separated by lines, similar to the usual honeycomb graph. However, a phase is now described by the occupancy of all four dots. (The overline denotes negative sign in the figure, e.g., $\bar{1}$=-1.) The left two numbers belong to the left DD, and the right two belong to the right DD. We denote the occupancy by $[N_1N_2;N_3N_4]$ where $N_i$ is the occupancy of the dot $D_i$. Note, that for the phase around $V_{left}=V_{right}=0$ we choose the [01;01] occupancy of our reference instead of [00;00]. It corresponds to simply a rigid shift of the operating point. In Fig. 3(a) the two DDs are independent of each other. By increasing the $V_{left}$ ($V_{right}$), only the occupancy of the left DD (right DD) changes. The occupancy of one dot of the DD increases by one, the other dot's occupancy decreases by one.

Fig. 3(b) shows the phase diagram for non-zero coupling between the DD's. The points where four phase borders meet are now split into two triple points. The square-shaped phase regions turn into hexagons. In Fig. 3(b) the crucial region of the phase diagram, which we examine experimentally, is framed. There are four phases in this region: [01;01], [01;10], [10;01] and [10;10]. During QCA operation the $V_{right}$ voltage is kept constant and $V_{left}$ changes sign. The system moves on a horizontal line in the phase diagram (shown by the arrow). By choosing an appropriate $V_{right}$, this horizontal line will cross the phase border between the [10;01] and [01;10] phases, corresponding to a transition from one polarization state to the other.

Figs. 4(a) and (b) show the phase borders where the left DD and the right DD, respectively, conduct. The experimental results of the conductance measurement corresponding to the framed parts of Figs. 4(a) and (b) are shown in Figs. 5(a) and (b). When only one DD conducts, the height of the conductance peak at the border is almost independent of the applied input voltages. However, at the phase borders, where both DD conduct, the conductance is significantly (up to 35-40%) decreased. The conductance lowering in the left and right DDs is clearly visible in the center of the corresponding conductance graphs of Figs. 5(a) and (b). The conductance lowering can be also seen in Fig. 6, where the conductance of the right DD is given as a function of $V_{right}$ for three different $V_{left}$ voltages. It is this lowering which the theoretical analysis of the next two sections will explain.

## III. Theory

We analyze the near-equilibrium behavior of circuits described in terms of *leads* and *metal islands* coupled by tunnel junctions and capacitors. The tunneling resistance is high enough ($R_t = 430 k\Omega \gg R_Q = h/e^2 \approx 26 k\Omega$, where $h$ is the Planck constant) to apply the perturbative theory. In modeling tunneling events the *orthodox theory*[1-4] of single electron tunneling was used, and *co-tunneling*[6-7,15-16] has been neglected, since the thermally assisted sequential tunneling plays the main role in conducting current.[41]

The free energy of a charge configuration can be written as

$$F = \frac{1}{2}\begin{bmatrix} q \\ q' \end{bmatrix}^T \mathbf{C}^{-1} \begin{bmatrix} q \\ q' \end{bmatrix} - v^T q' = E_{electrostatic} - W_{sources}. \qquad (1)$$

Here $\mathbf{C}$ is the capacitance matrix that describes the structure of the circuit, $v$ is a vector of lead voltages, and $q$ and $q'$ are the island charge vector and the lead charge vector, respectively. A metal island (dot) is connected to the circuit through capacitors and tunnel junctions, and its total charge is constrained to be (at $T=0K$) a multiple of the elementary charge.

The first term of the energy expression describes the electrostatic energy of the capacitors and tunnel junctions. The second term is the work done by the sources transferring charge to the leads. The equilibrium charge configuration for temperature $T=0$ K is the one that has minimal free energy. For $T>0$ K, higher energy configurations must also be included in computing thermal expectation values. The measured island charge is then no longer strictly an integer multiple of the elementary charge; it is rather the thermal average of the island charge over accessible configurations.

The dynamics of the system are governed by the following equation which gives the tunneling rate of an electron in a tunnel junction:[5]

$$\Gamma_{i \to j} = \frac{1}{e^2 R_T} \times \frac{\Delta F_{ij}}{1 - e^{-\frac{\Delta F_{ij}}{kT}}}, \qquad (2)$$

where $\Delta F_{ij}$ is the difference between the free energy of the initial and final states, and $R_T$ is the tunneling resistance of the junction. In most cases the change in free energy equals the difference of the free energies of the initial and final charge configurations ($\Delta F_{ij}=F_i-F_j$), except for the transitions when the electron enters to or arrives from a voltage source. In these cases $\Delta F_{ij}=F_i-F_j\pm eV_D$, where $V_D$ is the source voltage. The energy $eV_D$ is the work done by the voltage source to raise the potential of an elementary charge from ground to $V_D$.

The tunneling rates are used in a *master-equation*[1-3,17] model. An alternative approach would be the Monte Carlo method.[24-25] The master equation method is preferable here since the system is near equilibrium so the number of states (charge configurations) required for modeling is not large. For the master equation model, the accessible charge configurations and the transition rates between them must be known. Our model involves all the 16 charge configurations having 0 or 1 dot occupancies ([00;00], [00;01], [00;10],..., [11;10], [11;11]) and all the possible transitions connecting them.

The master equation has the form

$$\frac{dP}{dt} = \Gamma P, \qquad (3)$$

where $P$ is the vector containing the probabilities of occurrence for the 16 states and $\Gamma$ is a matrix describing the state transitions. This equation can be easily solved for the stationary state.

If the $V_{Dleft}$ and $V_{Dright}$ source voltages are zero then the $P_{st,i}$ stationary solutions are given by the Boltzmann distribution:

$$P_{st,i} = \frac{e^{-\frac{F_i}{kT}}}{\sum_k e^{-\frac{F_k}{kT}}}, \quad (4)$$

where $F_i$ is the free energy of state $i$. In this case the current is of course zero.

If the source voltages are small (experimentally they were 5 µV) then the $P_{st,i}$ stationary solutions can be approximated with the probabilities given by the Boltzmann distribution. The results are similar to those obtained from the master-equations. However, the Boltzmann distribution cannot be used to compute the current which is an inherently non-equilibrium phenomenon. Therefore the master equation approach is necessary for conductance computations.

Knowing the probability of occurrence for each state and the transition probabilities, the current through a hypothetical current meter can be computed as

$$I = e \sum_{i,j} c_{ij} P_{st,i} \Gamma_{i \to j}, \quad (5)$$

where $e$ is the elementary charge, $\Gamma_{i \to j}$ is the transition rate from state $i$ to state $j$, and $P_{st,i}$ is the $i^{th}$ element of stationary solution of the (3) master equations. The coefficient $c_{ij}$ is zero if the transition from state $i$ to state $j$ does not involve current through the current meter, and it is +1 (-1) if during this transition an electron exits (enters) through the current meter.

The master equation approach can also be used to compute the average transition rate between two charge configurations, even if there is not a direct transition between them. For example the transition time from state $i$ to state $j$ $(i<j)$ can be given in closed form as

$$\langle t_{i \to j} \rangle = r''_j \tilde{\Gamma}^{-2} \begin{bmatrix} 0 & \ldots & 0 & \overset{i}{\smile} & 0 & \ldots & 0 \end{bmatrix}^T, \qquad (6)$$

where the matrix $\tilde{\Gamma}$ and the row vector $r''_j$ are related to $\Gamma$. $\tilde{\Gamma}$ is obtained from $\Gamma$ omitting its $j^{th}$ row and $j^{th}$ column. $r''_j$ is obtained from the $j^{th}$ the row of $\Gamma$, leaving out its $j^{th}$ element. (For further details see the Appendix.)

## IV. Results and discussion

Based on the numerical solution of the master equations, Figs. 5(c) and (d) show the calculated conductances of the left and right double-dots as the functions of $V_{left}$ and $V_{right}$. (Compare with the experimental graphs shown in Figs. 5(a) and (b).) In Fig. 6 the computed conductance (solid line) and the measured conductance (crosses and dots) curves are shown as the function of $V_{right}$ for three different $V_{left}$ voltages. For the temperature the measured $T=70$ mK was taken.[39] Due to the unknown background charge, the conductance curve was allowed to shift rigidly in the $V_{left}$ and $V_{right}$ plane for fitting. The model uses the tunneling resistance as fitting parameter. The results of the calculations agree with the experiment upon taking $R_T=430$ kΩ. (The measured room temperature resistance of the tunnel junctions varied between 400 and 550 kΩ.) It can be observed that the conductance is lower on the phase border where both DDs conduct (in the center of the graphs in Figs. 5(c) and (d)), which matches the experiments.

We have shown that the solution of the master-equations for the two-DD system quantitatively agrees with the measured data. The master-equation model describes the correlated electron transport through the two DDs. This statement can be supported by computing the correlation between the charge polarization of the two DDs. The charge polarization of a DD is defined with the occupancy of the top and bottom dots as

$$P_{DD} = N_{top} - N_{bottom}. \qquad (7)$$

It is +1 and -1 for the [10] and the [01] double-dot charge configurations, respectively. We define the correlation function between the double-dots as:

$$C_{pp} = \langle P_{left} P_{right} \rangle - \langle P_{left} \rangle \langle P_{right} \rangle, \qquad (8)$$

where $\langle ... \rangle$ denotes the thermal expectation value. This correlation function would be zero if each DD only responded to the average charge on the other. In Fig. 7 the dependence of the correlation function is shown on the input voltages. $C_{pp}$ has a peak at the origin, where the conductance lowering occurs. Further from the origin its value is zero, indicating that there is no correlation between the double-dots there. The inset shows the temperature dependence of the correlation peak. The correlation between the double-dots decreases with increasing temperature. At the experimental temperature, the height of the correlation peak is $|C_{pp}| \cong 0.75$.

Correlated electron transport through the two DDs means that one DD responds to the instantaneous electron position in the other DD. It is instructive to examine what would happen if one DD responded only to the *average* charge density of the other DD. Fig. 8 shows the calculated conductance of the right DD in this case. (See Fig. 5(d) for comparison.) The conductance of the right DD was computed placing static charge in the left DD, corresponding to its time averaged charge density. The conductance lowering cannot be seen, and this also implies that the electron transport through the two DD is correlated.

In Fig. 5(d) $\Delta V_{right}$ denotes the voltage shift in the conductance graph of the right DD due to the change of occupancy in the left DD. If the coupling capacitance is higher between the two double-dots, this voltage shift and the conductance lowering will be larger.[40] However, if the two double-dots are coupled with smaller capacitances, $\Delta V_{right}$ and the conductance lowering decreases. In the limit of uncoupled DDs, conductance lowering does not occur and $\Delta V_{right}=0$.

We can use this analysis to estimate the P=+1/P=-1 transition rate. The results of the computations give 50MHz for this particular two-DD structure. During the P=+1/P=-1 transition the input voltage of the left DD is changed, while the input voltage of the right DD is kept constant. The input voltage of the left DD is changed in such a way that it mimics the switching of an adjacent cell.[37] Modifying the capacitances, especially the coupling between the two DDs, and decreasing the resistance of the tunnel junctions can increase the transition frequency.[40]

## V. Conclusions

In this paper electron transport through coupled double-dots has been analyzed. Experimentally, a suppression of conductance in one double-dot was observed when the second double-dot was conducting. This is explained theoretically in terms of the correlation of electron motion in the system. A model has been developed which rather accurately reproduced the experimental data. The straightforward interpretation of this model is that the electron in one double-dot responds not just to the time average fluctuations of charge in the neighboring double-dot, but to the instantaneous charge configuration. This leads to a non-vanishing correlation in the coupled electron motion.

## Acknowledgment

We are very grateful to John Timler, István Daruka and Ikaros Bigi for valuable conversations. This work was supported by DARPA and NSF.

## Appendix: Average transition time computations

The average transition time from state $i$ to state $j$ ($i<j$) is computed. The computations are based on the following model. First all the systems of the ensemble are in state $i$. Then the ensemble is allowed to evolve according to the master equation describing its behavior. (It will be given later.) Eventually all the systems arrive at state $j$ ($P_j(\infty)=1$). The average transition time can be computed as

$$\langle t_{i \to j} \rangle = \int_0^\infty t \frac{dP_j}{dt} dt, \qquad (9)$$

where $\frac{dP_j}{dt} \times \Delta t$ gives the ratio of systems which reach state $j$ during the $\Delta t$ time interval.

When measuring transition time from state $i$ to state $j$ the systems already arrived in state $j$ should stay in state $j$ and should not leave it. Thus the $\Gamma$' coefficient matrix used for average transition time computations is different from the original $\Gamma$ matrix of the system. It can be obtained from $\Gamma$ by setting the elements of its $j^{th}$ column to zero. (This corresponds to the inhibition of all the transitions from state $j$.) The master equation with the modified $\Gamma$' coefficient matrix is:

$$\frac{dP}{dt} = \Gamma' P. \qquad (10)$$

The $P(t)$ solution of this equation can be written in an exponential form. From this solution the $\frac{dP_j}{dt}$ can be expressed and substituted into (9); however, the integration cannot be done symbolically because $\Gamma$' is not invertible. (To compute the integral given in (9) we need the inverse of $\Gamma$'.) Thus, before making the steps just mentioned, some additional matrix manipulations are needed to make $\Gamma$' invertible.

One way to make $\Gamma$' invertible is to eliminate $P_j$ from (10). $P_j$ can be easily eliminated because the $j^{th}$ column of the coefficient matrix is only zeros. The elimination of $P_j$ corresponds to changes in the coefficient matrix and the $P$ vector. The new coefficient matrix, $\tilde{\Gamma}$, is obtained from $\Gamma$' omitting its $j^{th}$ row and $j^{th}$ column. It can be obtained from $\Gamma$' as well with the same transformation, because $\Gamma$' and $\tilde{\Gamma}$ differ only in $j^{th}$ the column that was just omitted. $\tilde{P}$ is formed by leaving out the $j^{th}$ element of $P$.

After the elimination of $P_j$ the following master equation is obtained:

$$\frac{d\tilde{P}}{dt} = \tilde{\Gamma}\tilde{P}. \tag{11}$$

The initial value of $\tilde{P}$ corresponds to the case when all the systems of the ensemble are in state $i$:

$$\tilde{P}(0) = \begin{bmatrix} 0 & \ldots & 0 & \overset{i}{1} & 0 & \ldots & 0 \end{bmatrix}^T. \tag{12}$$

The time dependence of $\tilde{P}$ can be given as the solution of the (11) master equation:

$$\tilde{P}(t) = e^{\tilde{\Gamma}t}\tilde{P}(0) = e^{\tilde{\Gamma}t}\begin{bmatrix} 0 & \ldots & 0 & \overset{i}{1} & 0 & \ldots & 0 \end{bmatrix}^T. \tag{13}$$

For (9) we need the time derivative of $P_j$, but (11) does not contain it because it was obtained after eliminating $P_j$. The time derivative of $P_j$ can be found in (10). This master equation represents a differential equation system. The $j^{th}$ line of the equation system that gives the required time derivative is:

$$\frac{dP_j}{dt} = r'_j P, \tag{14}$$

where $r'_j$ is the $j^{th}$ row of $\Gamma$'. Knowing that the $j^{th}$ element of $r'_j$ is zero (the $j^{th}$ column of $\Gamma$' is zero) this can be written with $\tilde{P}$ as

$$\frac{dP_j}{dt} = r''_j \tilde{P}, \tag{15}$$

where $r''_j$ is the $j^{th}$ row of $\Gamma$' (and also of $\Gamma$) omitting its $j^{th}$ element. Substituting first (15) and then (13) into (9), the average transition time from state $i$ to state $j$ is:

$$\langle t_{i \to j} \rangle = \int_0^\infty t r''_j \tilde{P} dt = r''_j \int_0^\infty t \tilde{P} dt = r''_j \int_0^\infty t e^{\tilde{\Gamma}t} dt \begin{bmatrix} 0 & \ldots & 0 & 1 & 0 & \ldots & 0 \end{bmatrix}^T. \tag{16}$$

Using

$$\int_0^\infty t e^{-\alpha t} dt = \alpha^{-2}, \qquad (17)$$

the transition time in a closed form is obtained as (6). The right hand side of (17) can be computed because $\tilde{\Gamma}$ is invertible. The infinite integral can be evaluated because $\tilde{\Gamma}$ has only negative eigenvalues.

[40] For a subsequent experimental setup the conductance lowering reached 50%. The computed transition frequency was 200 MHz.

[41] One reason for that is the high tunneling resistance. The co-tunneling rate is proportional to $(R_t)^{-2}$ while the sequential tunneling rate is proportional to $(R_t)^{-1}$ thus large tunneling resistance helps suppressing co-tunneling. Another reason is the low source voltage ($eV_{Dleft}, eV_{Dright} < kT = 6.1 \mu eV$). The co-tunneling rate is proportional to the third power of $V$ thus for high enough voltages it could be dominant. Finally, although to get current through the DD, tunneling to higher energy states is necessary, these transitions are never suppressed enough to make co-tunneling important (at least in the cases when considerable current flows). Based on co-tunneling rate computations,[3] the error due to neglecting co-tunneling turns out to be always less than 5% of the maximum conductance.

## Figure Captions

FIGURE 1. Schematic of the basic four-site semiconductor QCA cell. (a) The geometry of the cell. The lines indicate the possibility of interdot tunneling. The tunneling energy between two sites (quantum dots) is determined by the heights of the potential barrier between them. (b) Coulombic repulsion causes the two electrons to occupy antipodal sites within the cell. These two bistable states result in cell polarization of P=+1 and P=-1.

FIGURE 2. (a) Two-DD system. The $D_1$, $D_2$, $D_3$ and $D_4$ denote the four metal islands (dots). The $V_{Dleft}/V_{Dright}$ voltage sources and the $I_{left}/I_{right}$ current meters are used for double-dot conductance measurements. (b) The symbolic representation of the system. The circles and the lines represent metal islands and tunnel junctions, respectively.

FIGURE 3. (a) The phase diagram of the two-DD system if there is no capacitive coupling between the left and right DDs. The figure shows the $[N_1N_2;N_3N_4]$ most probable charge configuration as the function of the input voltages. (b) The phase diagram of the two-DD system when the left and right DDs are capacitively coupled. The framed part of the phase diagram is studied in this paper. At the phase borders one of the DDs (e.g., [01;10]/[01;01]) or both of them (e.g., [01;10]/[10;01]) conduct. The arrow corresponds to QCA operation.

FIGURE 4. The phase borders where the (a) left and the (b) right DD conduct. The conductances for the framed part are shown in Fig. 5 magnified.

FIGURE 5. Comparison of the (a-b) measured and the (c-d) calculated conductance curves of the left and right double-dots. The conductances are given as a function of $V_{left}$ and $V_{right}$. In (d) the $\Delta V_{right}$ voltage shift is the effect of the change of occupancy in the left DD. The 10, 20, 30, 40 and 50 nS contours are shown. The conductance suppression is clearly visible in the center of the graphs. For (c) and (d) the insets show

the three-dimensional conductance plots. The curves corresponding to the three vertical lines in (b) are given in Fig. 6.

FIGURE 6. The measured (crosses and dots) and computed (solid line) conductance curves as the function of $V_{right}$ for three different $V_{left}$ voltages. The curves correspond to the three vertical lines in Fig 5(b).

FIGURE 7. The correlation between the top dots of the two DDs as a function of $V_{left}$ and $V_{right}$ for $T$=70 mK. The correlation is maximum at the origin where the conductance lowering occurs. The inset shows the temperature dependence of the correlation peak. It decreases with increasing temperature.

FIGURE 8. The calculated conductance of the right DD for the case if the right DD responded to the average charges on the left DD. In the graph, the 10, 20, 30, 40 and 50 nS contours are shown. The conductance lowering is not seen in this figure. (Compare with Fig. 5(d).)

**Figures**

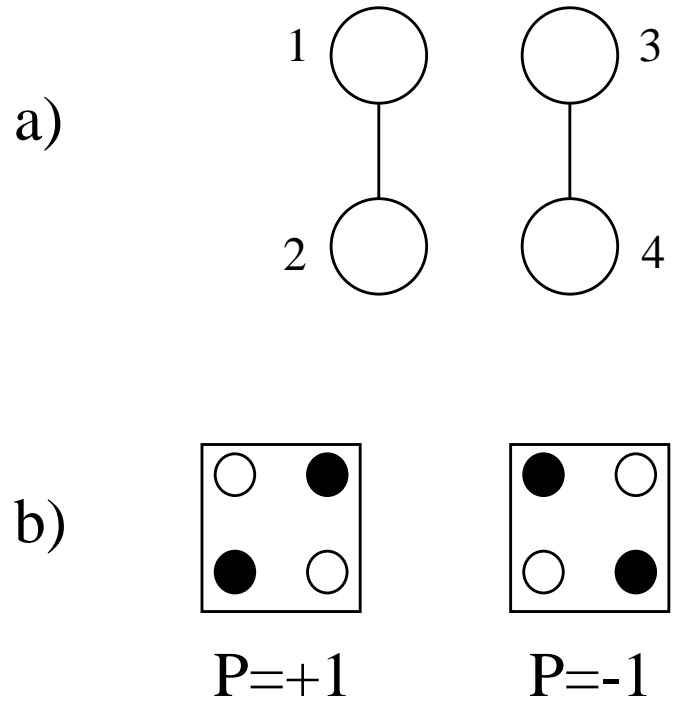

Figure 1

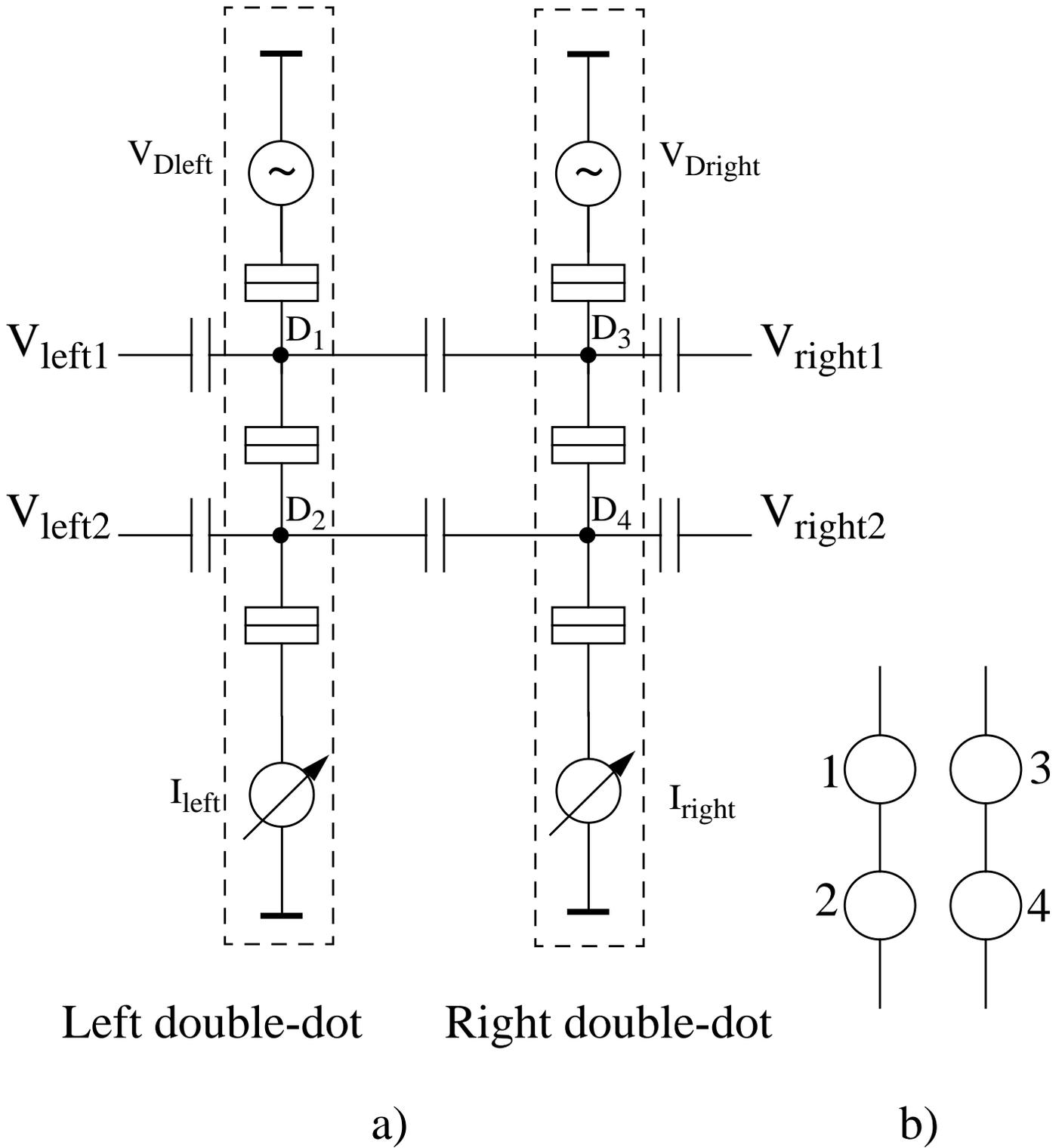

Left double-dot     Right double-dot

a)           b)

Figure 2

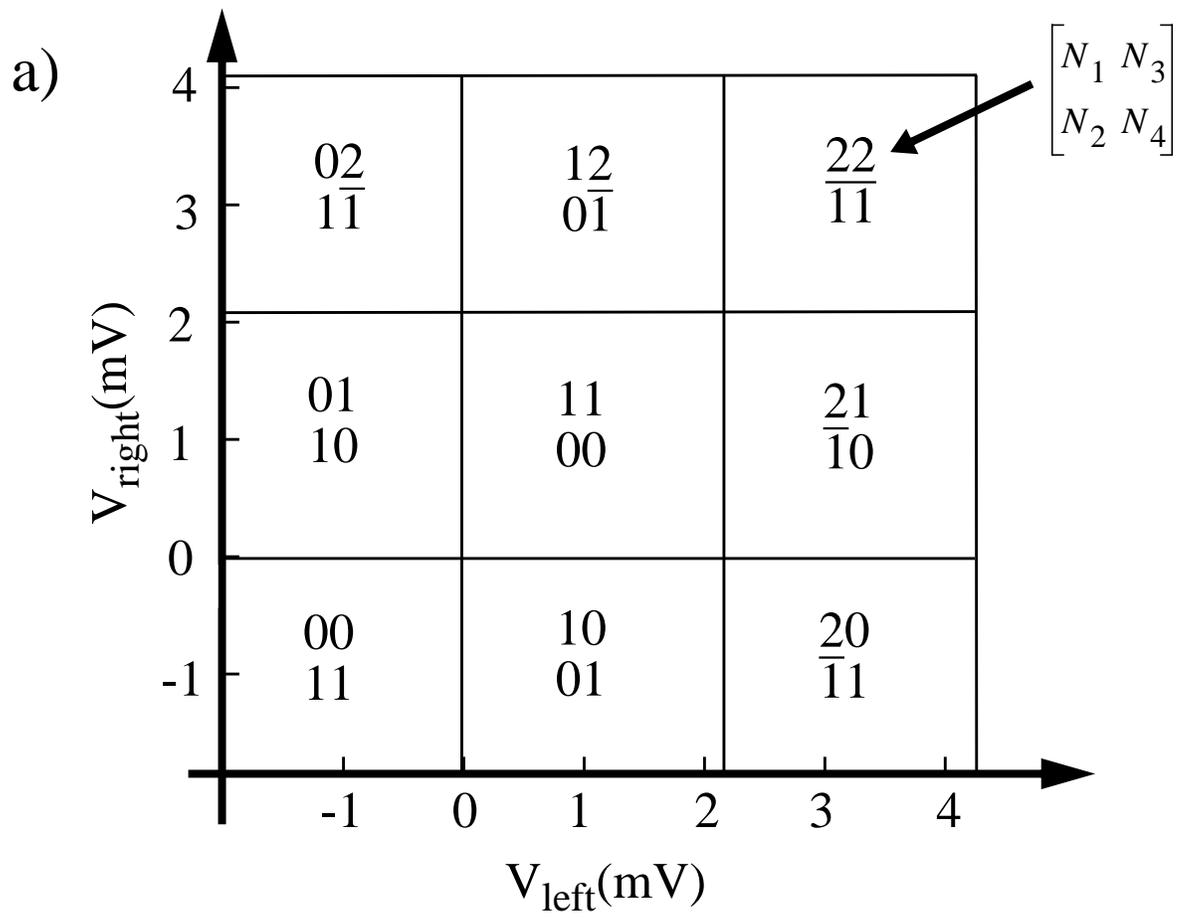

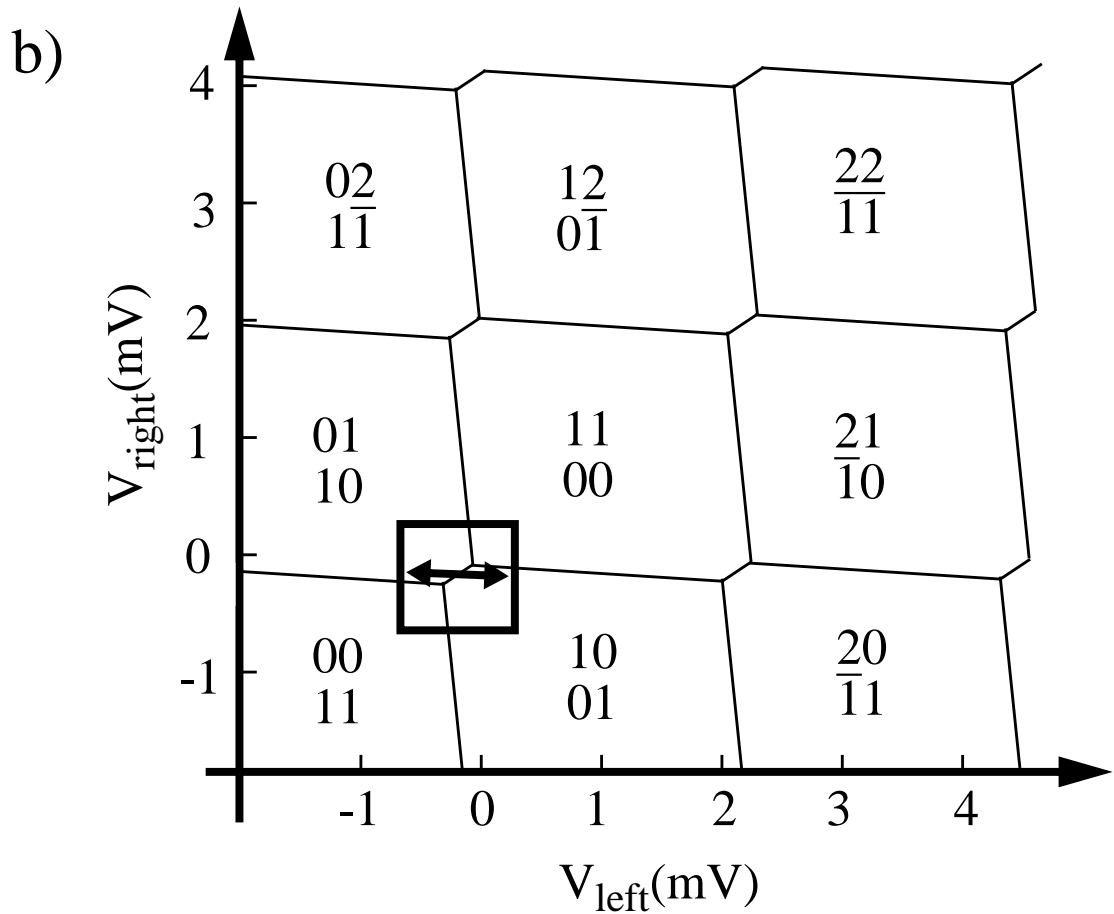

Figure 3

a)

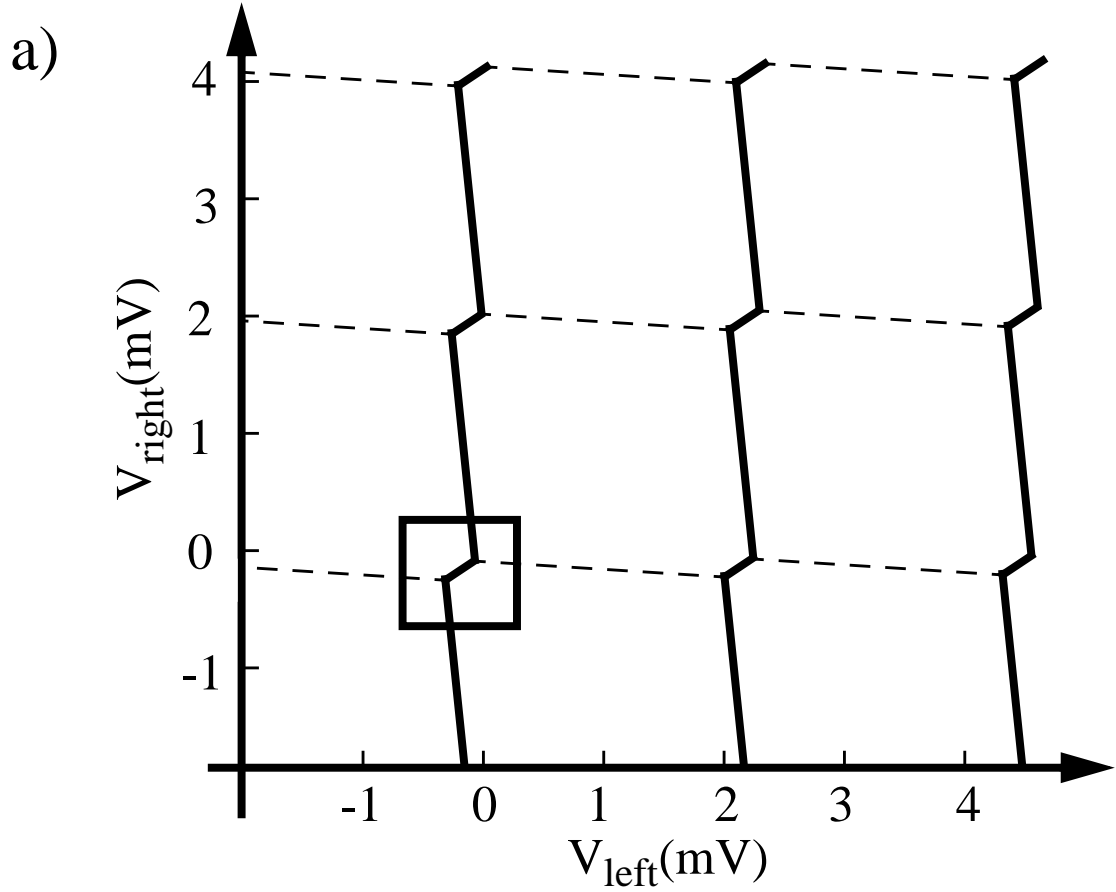

b)

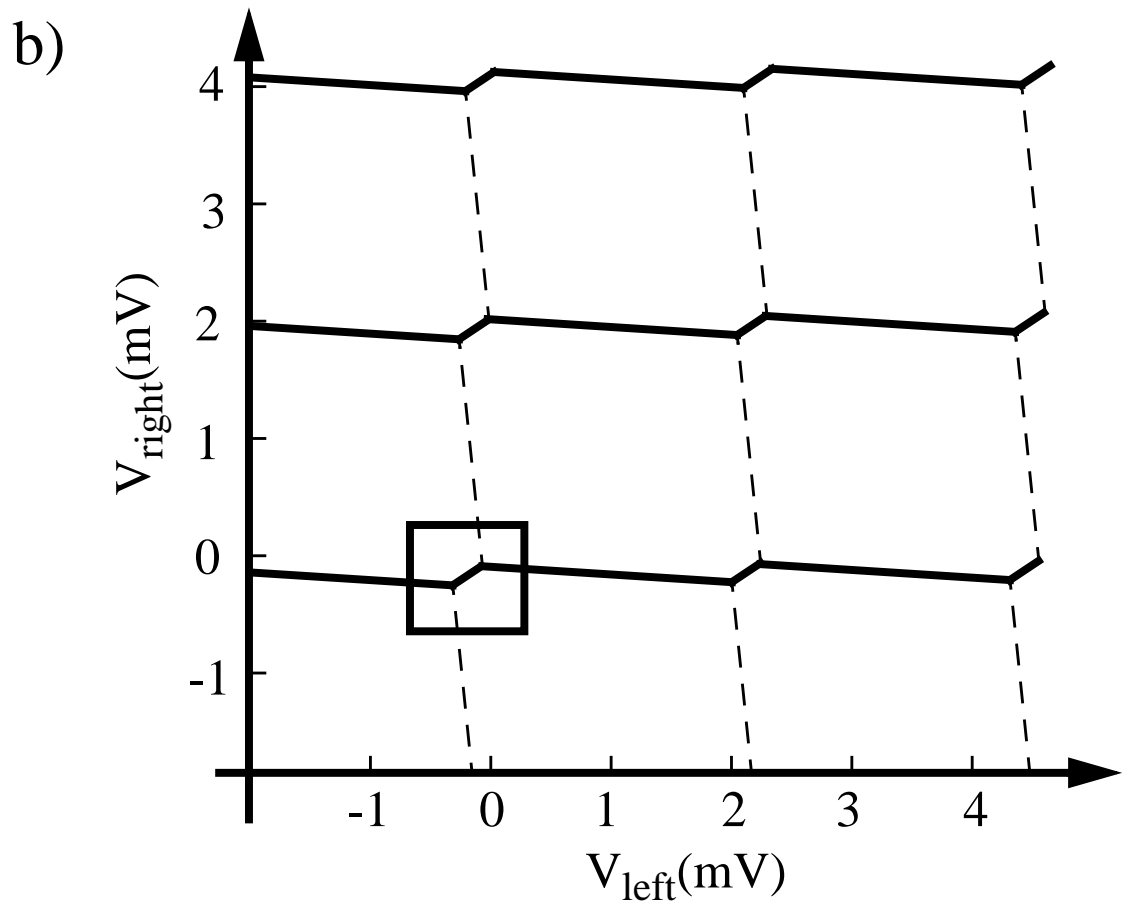

Figure 4

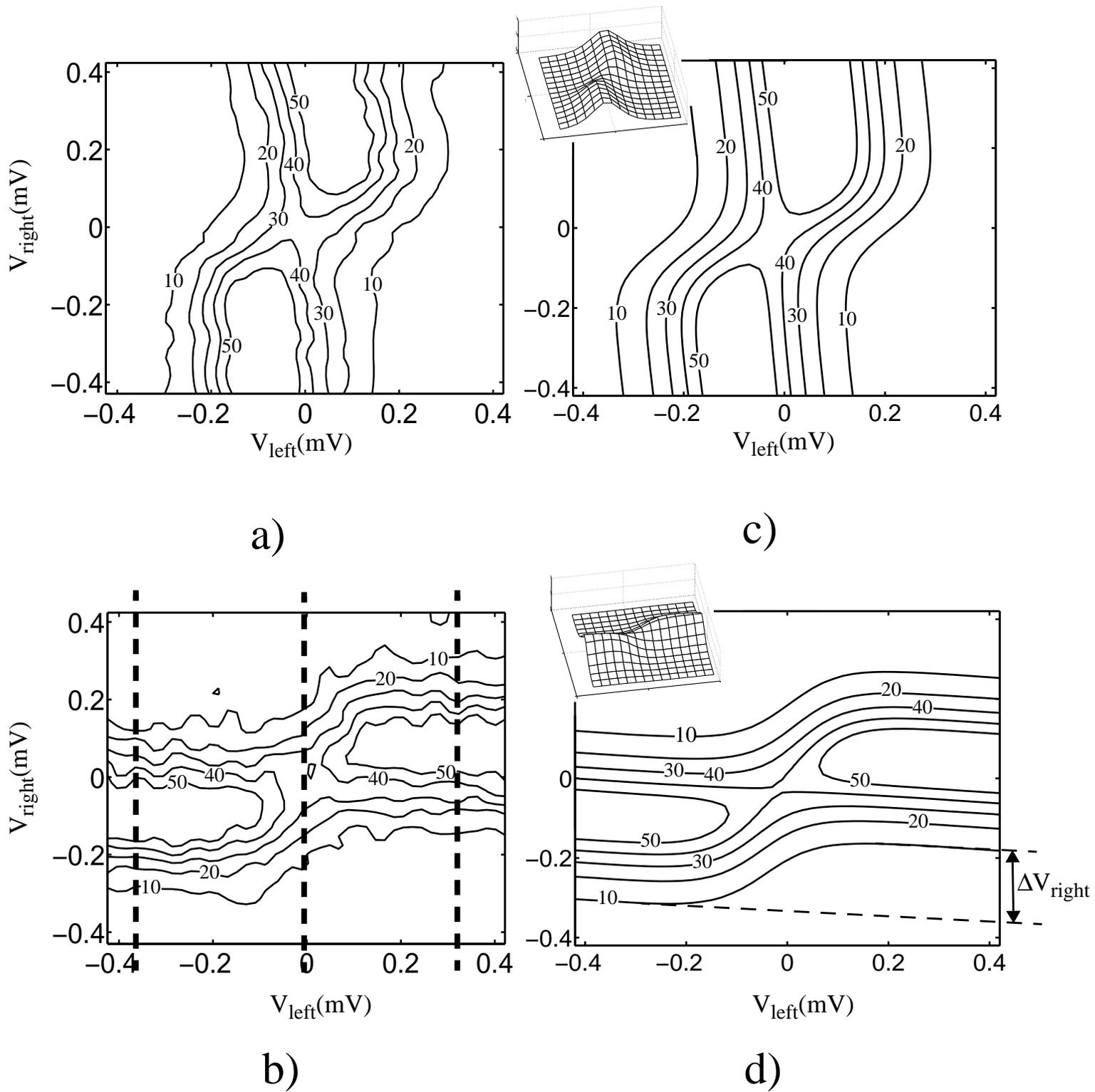

Figure 5

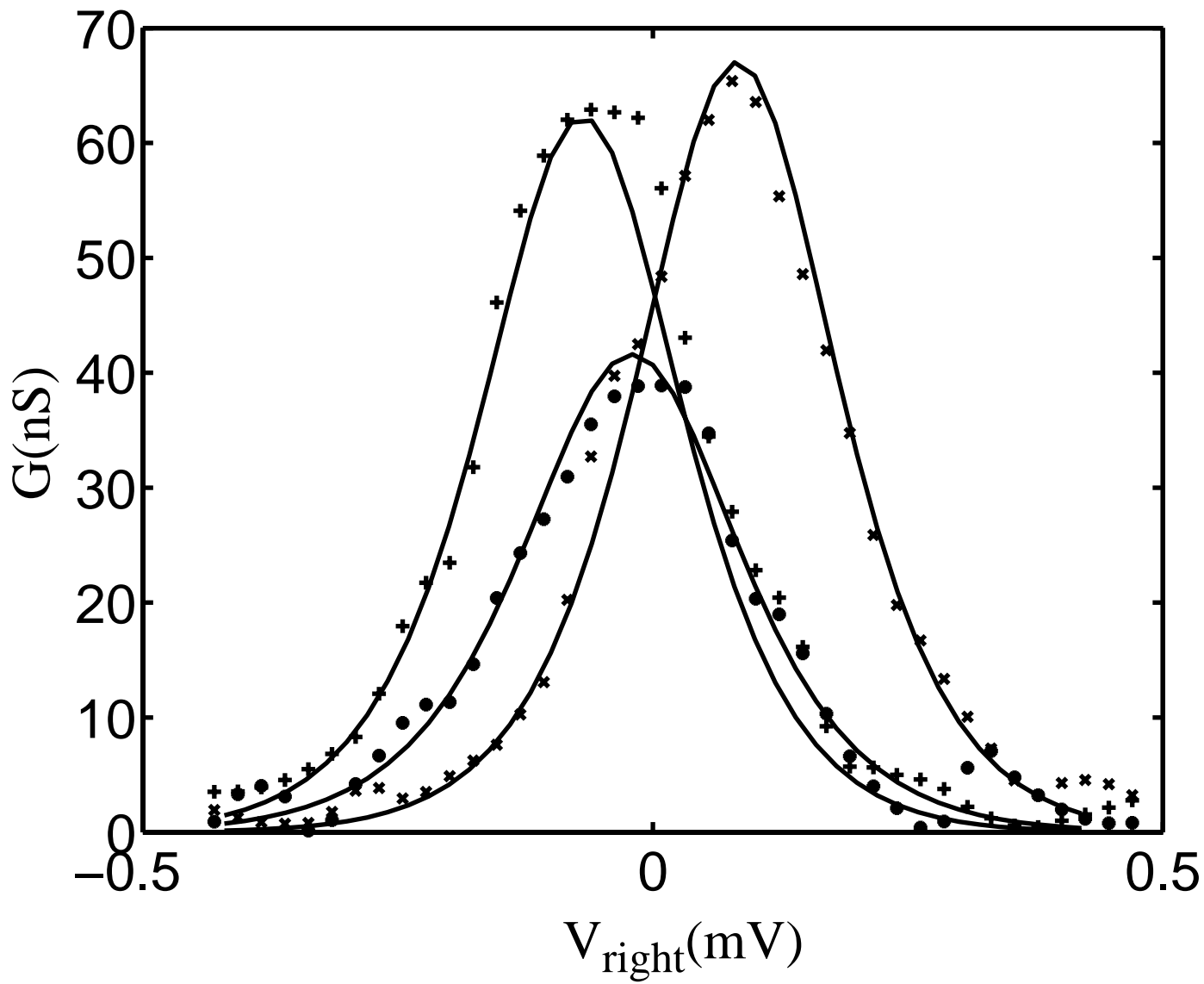

Figure 6

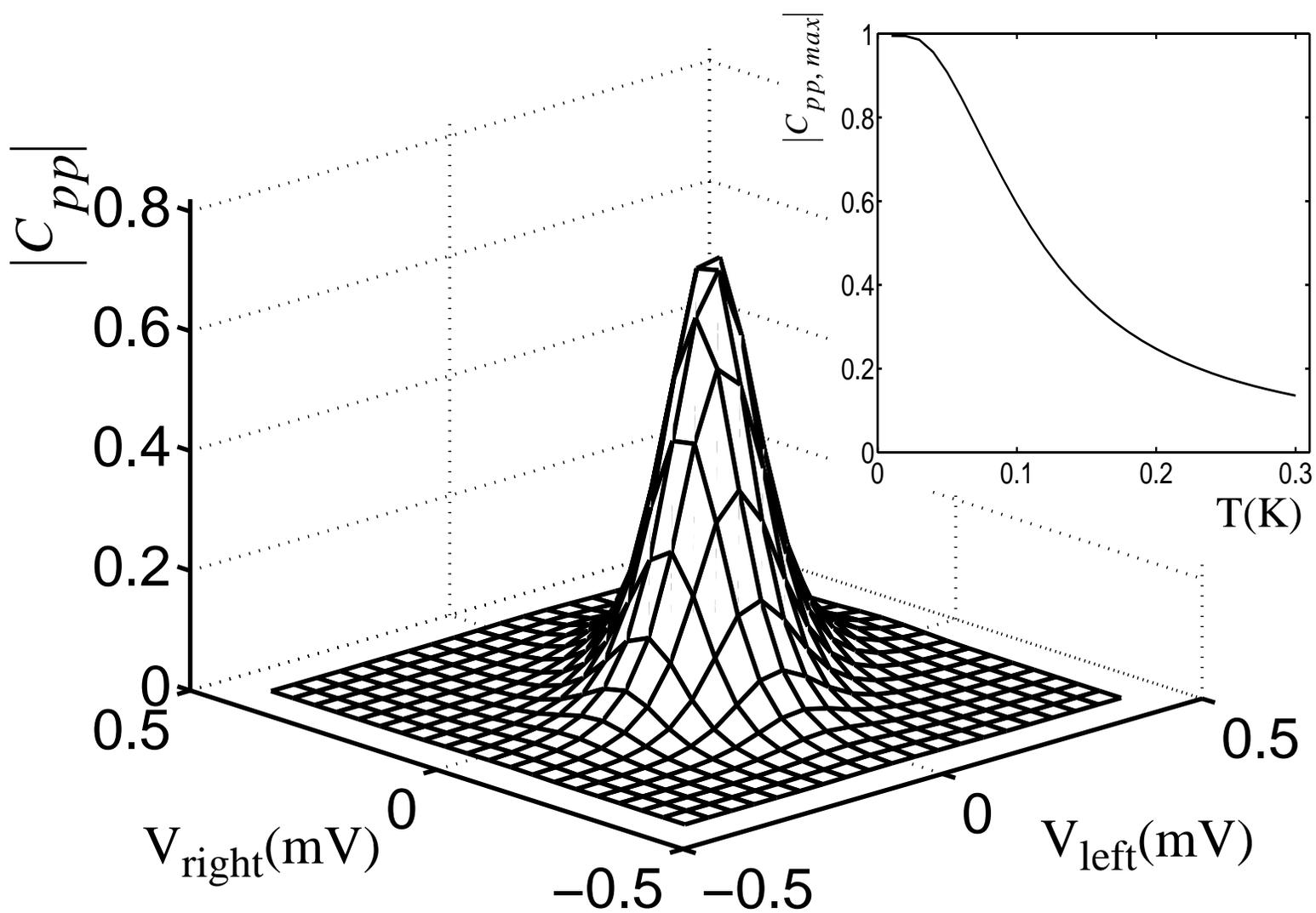

Figure 7

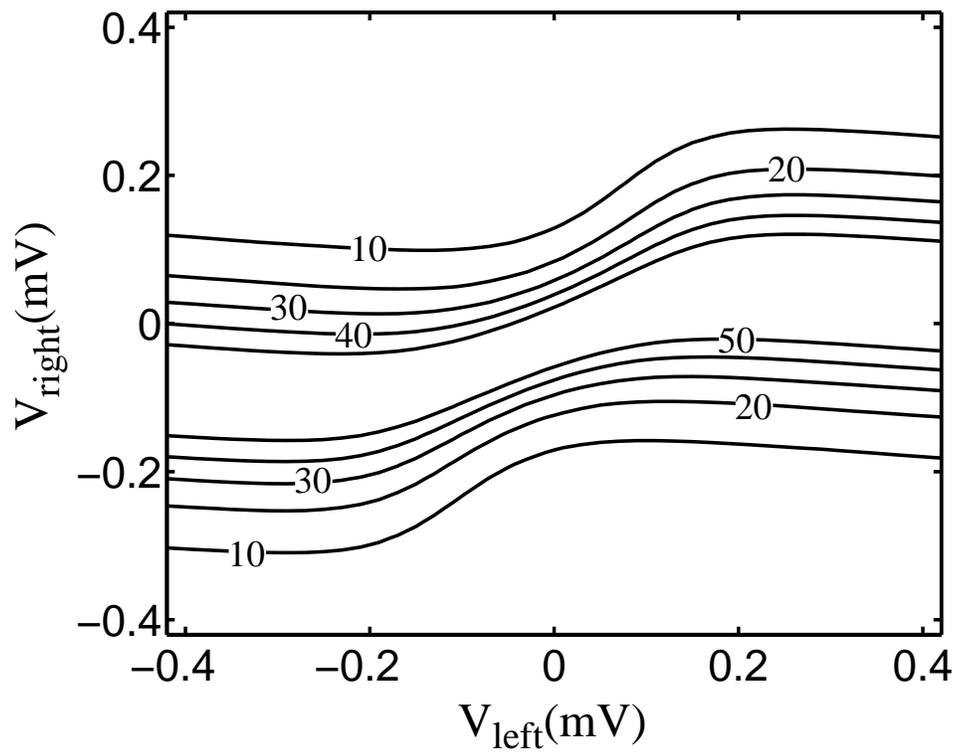

Figure 8